\documentclass[prl,twocolumn,showpacs,preprintnumbers,amsmath,amssymb,citeautoscript]{revtex4-1}

\usepackage{graphicx}
\usepackage{amssymb, amsfonts, amsmath}
\usepackage{dcolumn}
\usepackage{bm}
\usepackage{color}

\newlength{\figwidth}
\begin{document}

\title{Strong suppression of superconductivity by divalent Ytterbium Kondo-holes in CeCoIn$_5$}

\author{M.~Shimozawa$^1$}
\author{T.~Watashige$^1$}
\author{S.~Yasumoto$^1$}
\author{Y.~Mizukami$^1$}
\author{M.~Nakamura$^2$}
\author{H.~Shishido$^3$}
\author{S.~K.~Goh$^1$}
\author{T.~Terashima$^2$}
\author{T.~Shibauchi$^1$}
\author{Y.~Matsuda$^1$}

\affiliation{
$^1$Department of Physics, Kyoto University, Kyoto 606-8502, Japan\\
$^2$Research Center for Low Temperature and Materials Sciences, Kyoto University, Kyoto 606-8501, Japan\\
$^3$Department of Physics and Electronics, Osaka Prefecture University, Osaka, Japan
}

\date{\today}

\begin{abstract}

To study the nature of partially substituted Yb-ions in a Ce-based Kondo lattice, we fabricated high quality Ce$_{1-x}$Yb$_x$CoIn$_5$ epitaxial thin films using molecular beam epitaxy.  We find that the Yb-substitution leads to a linear decrease of the unit cell volume, indicating that Yb-ions are divalent forming Kondo-holes in  Ce$_{1-x}$Yb$_x$CoIn$_5$,  and  leads to a strong suppression of  the superconductivity and Kondo coherence.   These results, combined with the measurements of  Hall effect,  indicate that  Yb-ions act as nonmagnetic impurity scatters in the coherent Kondo lattice without serious suppression of the antiferromagnetic fluctuations.   These are in stark contrast to previous studies performed using bulk single crystals, which claim the importance of valence fluctuations of Yb-ions.    The present work also highlights the suitability of epitaxial films in the study of the impurity effect on the Kondo lattice.

\end{abstract}


\maketitle

The impurity effect has been one of the key issues for the study of unconventional superconductors  including cuprates \cite{Bal06}, iron-pnictides \cite{Ste11} and heavy fermion compounds \cite{Pag07,Cap210,Bau11,Shu11,Cap10,Boo11,Ham11}.  In these strongly correlated electron systems, the superconducting state emerges from the competition between different phases, such as a magnetically ordered phase.  Understanding the  impurity effect is important because it  is intimately related to the superconducting pairing interaction and  competing electronic correlations.  In heavy fermion compounds, it was generally found that  a substitution of magnetic sites with nonmagnetic impurities yields a substantial reduction in both the onset of  coherence temperature $T_{\rm coh}$ at which the formation of  heavy fermions occurs, and the superconducting transition temperature $T_c$.   Recently, however, highly unusual impurity effect has been reported in the  heavy fermion compound CeCoIn$_5$, which is in sharp contrast to the above tendency \cite{Shu11,Cap10,Boo11}.

 CeCoIn$_5$ with tetragonal symmetry \cite{Pet01} is a very clean superconductor with large mean free path, which is known to exhibit a plethora of fascinating electronic properties.  The unconventional superconductivity with $d$-wave symmetry appears in the vicinity of magnetic ordered phase \cite{Izawa01, An10, Sto08}.    The normal state exhibits pronounced non-Fermi-liquid behaviors, which are believed to be due to the proximity of an antiferromagnetic (AF) quantum critical point \cite{Sid02, Nakajima07}.    Recent study revealed that the superconductivity occurs in two-dimensional Kondo lattice composed of square lattice of Ce atoms \cite{Mizukami11}.  Thus CeCoIn$_5$ may provide an ideal playground for investigating the impurity effect on unconventional superconductivity in strongly correlated $f$-electron systems.  The impurity effect on CeCoIn$_5$  has been investigated extensively by rare earth, $R$, (or Hg, Cd, Sn) substitution for Ce (or In) \cite{Pag07, Nak02, Ram10, Cap210}.  Two notable features have been pointed out.   Firstly,  non-magnetic impurities locally suppress superconductivity, generating an inhomogeneous electronic \textquotedblleft Swiss cheese\textquotedblright , similar to cuprates \cite{Bau11}.  Secondly, several groups have reported more striking and unexpected results in the case of Yb-substitution on Ce-site: the superconductivity  in Ce$_{1-x}$Yb$_x$CoIn$_5$ is robust against Yb-substitution and $T_c$ decreases linearly with $x$ towards 0~K as $x \rightarrow $~1, while $T_{\rm coh}$ remains essentially unaffected with Yb-doping \cite{Shu11, Cap10, Boo11}.  YbCoIn$_5$ is a conventional nonmagnetic metal with no  superconducting transition down to 20~mK, indicating that Yb is close to divalent and in the nonmagnetic closed shell 4$f^{14}$ configuration \cite{Zar03,Mizukami11}.  Therefore, the robustness of the superconductivity upon Yb-substitution in  CeCoIn$_5$ with unconventional pairing symmetry is extraordinary.   Moreover, according to Refs.\cite{Shu11} and \cite{Cap10}, the lattice parameters remain roughly constant as $x$ changes until the phase separation occurs at $x\sim$~0.8, indicating the violation of Vegard's law.   These results are markedly different from other $R$ substituted CeCoIn$_5$,  in which $R$ substitutions suppress the superconductivity at approximately $x$~=~0.2-0.3 \cite{Pet02,Nak02}.  These anomalous behaviors have been discussed in the light of the valence fluctuations of Yb-ions \cite{Shu11}.  However,  the electronic state of Yb in the Kondo lattice of CeCoIn$_5$ is poorly understood.  

To obtain further insight into the nature of Yb-ion in the Ce-based Kondo lattice, we fabricated  Ce$_{1-x}$Yb$_x$CoIn$_5$ thin films using molecular beam epitaxy (MBE) technique.  This technique, recently advanced by our group, enables the growth of uniform heavy fermion thin films,  as demonstrated by the successful growths of CeIn$_3$/LaIn$_3$ \cite{Shishido10} and superconducting CeCoIn$_5$/YbCoIn$_5$ \cite{Mizukami11} superlattices with one-unit-cell thick CeIn$_3$ and CeCoIn$_5$ layers, respectively. Since the epitaxial growth occurs in non equilibrium condition at temperatures much lower than the temperature used for single crystal growth by flux method,  MBE is suitable for the preparation of systems with homogeneously distributed Ce and Yb-atom.   Moreover, owing to the ability to evaporate Ce and Yb atoms simultaneously, the ratio of Ce to Yb can be controlled precisely.   Using our thin films, we find that the Yb-substitution leads to a strong suppression of $T_c$, in accordance with Abrikosov-Gorkov (AG) theory.  These results are in sharp contrast to previous studies performed using bulk single crystals \cite{Bau11,Shu11,Cap10,Boo11}.

\begin{figure}[t]
\begin{center}\leavevmode
\hspace*{-10mm}
\includegraphics[width=1.17\linewidth]{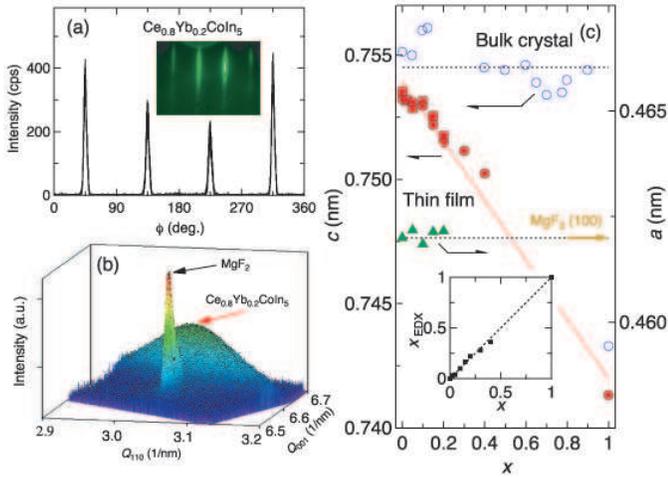}
\caption{(color online). (a) X-ray diffraction $\phi$-scan data of the (115) peak for Ce$_{0.8}$Yb$_{0.2}$CoIn$_5$. Inset : Streak patterns of the RHEED image during the crystal growth. (b) X-ray reciprocal lattice mapping for Ce$_{0.8}$Yb$_{0.2}$CoIn$_5$ near the (115) peak. (c) Lattice parameters $a$ (filled green triangles) and $c$ (filled red squares and  circles) for Ce$_{1-x}$Yb$_x$CoIn$_5$ films.  Here $c$ is determined by reciprocal mapping (squares) and $\theta-2\theta$ scan (circles).  Open blue circles represent $c$ of bulk crystals reported in Ref.\cite{Shu11}. The dased lines are guides for eyes.   The inset shows $x$ determined by EDX, $x_{\rm EDX}$, vs. $x$ determined by the evaporation ratio of Ce to Yb atoms.    }
\hspace*{10mm}
\end{center}
\end{figure}

 The $c$ axis oriented epitaxial Ce$_{1-x}$Yb$_x$CoIn$_5$  films  were grown by MBE.  The (001) surface of MgF$_2$ with rutile structure ($a$~=~0.462~nm, $c$~=~0.305~nm) was used as a substrate.  The substrate temperature was kept at 530-550~$^{\circ}$C, depending on the composition.   Each metal element was evaporated from individually controlled Knudsen-cell. The typical deposition rate was 0.01-0.02 nm/s. The area and thickness of the films were $5.0\times10.0$~mm$^2$ and 120~nm, respectively.    The epitaxial growth of each layer with atomic flatness was carefully checked  by monitoring the streak patterns of the Reflection High Energy Electron Diffraction (RHEED) during the deposition (the inset of Fig.~1a).   The X-ray diffraction $\phi$-scan with 4-fold peaks shown in Fig.~1a also indicates the epitaxial growth of the film, which was further confirmed by the X-ray reciprocal lattice mapping as shown in Fig~1b.    Surface flatness in atomic level was confirmed by the atomic force microscopy.   Excellent chemical homogeneity across the whole sample area was confirmed by the  energy dispersive X-ray spectroscopy (EDX) analysis.  We could not grow the epitaxial films for $x>0.4$ without buffer layers.    To grow YbCoIn$_5$ epitaxial  films, we used CeIn$_3$ as buffer layers on the substrate.     As shown in the inset of Fig.~1c, the ratio of  Ce to Yb of the measured films  determined with EDX coincides well with the evaporation ratio of Ce to Yb atoms.  

 In our epitaxial thin films ($a$~=~0.462~nm, $c$~=~0.753~nm), lattice parameter $a$ is dictated by the substrate lattice parameter and misfit strain slightly enlarges $a$ from that of  bulk single crystal value ($a$~=~0.461~nm, $c$~=~0.755~nm).   Figure~1c  shows the doping evolution of  lattice parameters determined by X-ray diffraction.  Because of the epitaxial strain effect,   $a$ is independent of $x$ and  coincides well with $a$ of MgF$_2$ substrate.    On the other hand,  $c$ (and hence unit cell volume)  decreases linearly with increasing $x$ and lies on the line connecting $x$~=~0 and 1.   According to the Vegard's law, the unit cell volume should decreases linearly with Yb-substitution, if  there are no changes in the valence of Ce and Yb ions.  Therefore the observed linear relation implies that the Ce and Yb ions in the present  Ce$_{1-x}$Yb$_x$CoIn$_5$ films retain the valences of the end member compounds, i.e. the Yb-ion is divalent in Ce$_{1-x}$Yb$_x$CoIn$_5$, forming nonmagnetic ``Kondo holes" in the 4$f^1$ lattice.    This result is in sharp contrast to the results using bulk single crystals, in which $c$ is $x$-independent (open circles in Fig.~1c) \cite{Shu11,Cap10}.  

\begin{figure}[t]
\hspace*{-3mm}
\includegraphics[width=1.05\linewidth]{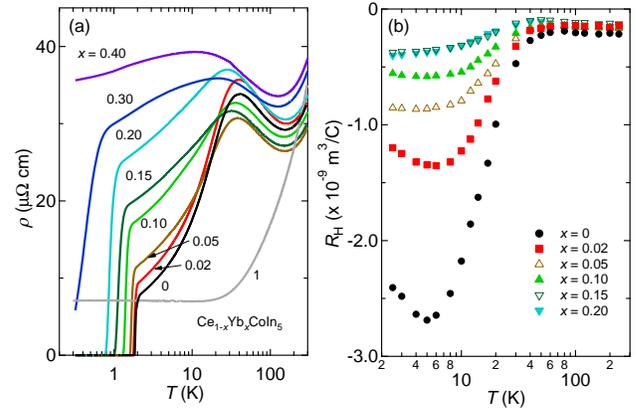}
\caption{(color online). Temperature dependence of the (a) resistivity  and  (b) Hall coefficient for  Ce$_{1-x}$Yb$_x$CoIn$_5$.}
\hspace*{3mm}
\end{figure}

 Figure~2a and 2b depict the temperature dependence of the in-plane resistivity $\rho$  and Hall coefficient $R_{\rm H}$ for {\boldmath $H$}$\parallel c$, which is defined as the field derivative of  Hall resistivity at the zero field limit, respectively.   
First we compare the transport  properties of pure CeCoIn$_5$ thin film with those of single crystal.  The temperature dependence of both the resistivity and Hall coefficient in thin film is essentially the same as those in bulk single crystals \cite{Nakajima07}.    Associated with an incoherent-coherent crossover, $\rho$ shows a maximum at $T_{\rm coh}$ at around 40~K.   The absolute value of $\rho$ of CeCoIn$_5$ thin film is  close to that of bulk single crystal.  In particular, $\rho\simeq 6~\mu \Omega$~cm at the superconducting onset is close to the bulk single crystal value, which ranges from 3 to 8~$\mu \Omega$~cm \cite{Pag07,Mov01,Nic01}.   The resistivity below $T_{\rm coh}$ in bulk single crystal exhibits a $T$-linear dependence, which is a hallmark of non-Fermi liquid behavior.  In our thin film, the resistivity exhibits a $T$-dependence with exponent slightly below unity. 

  $R_{\rm H}$ is negative in the whole temperature range.  The temperature dependence of $R_{\rm H}$  is closely correlated with that of the resistivity \cite{Nakajima07,Nakajima08}.      At high temperatures $T>T_{\rm coh}$, $R_{\rm H}$ is nearly temperature  independent.  Below $T_{\rm coh}$, $R_{\rm H}$ decreases rapidly as the temperature is lowered.  Further decrease of temperature increases $R_{\rm H}$ after showing a minimum at around 5~K.  The magnitude of  $R_{\rm H}$ above $T_{\rm coh}$ matches well with that of bulk single crystal \cite{Nakajima07}.   The enhancement of the absolute value of $R_{\rm H}$ below $T_{\rm coh}$ is nearly 15 times that at above $T_{\rm coh}$, which is nearly half of that in bulk single crystal.  The superconducting transition temperature of the CeCoIn$_5$ thin film is 1.95~K, which is slightly lower than $T_c=2.3~$K of bulk single crystal.  Since the $\rho$ value at the superconducting onset (and residual resistivity which will be discussed later) in thin film is comparable to the bulk single crystal value, this $T_c$ reduction is likely to be due to the strain effect arising from the substrate.  In fact, it has been reported that the negative pressure in CeCoIn$_5$ reduces $T_c$ \cite{Pha06}.   Based on these results, we conclude that the quality of our epitaxial thin films grown by MBE is comparable to that of high quality bulk single crystals. 

\begin{figure}[t]
\begin{center}\leavevmode
\hspace*{-3mm}
\includegraphics[width=1.07\linewidth]{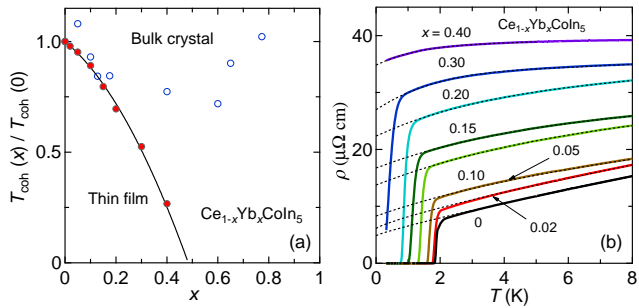}
\caption{(color online). (a) $x$-dependence of $T_{\rm coh}$ normalized by the value at $x$=0 for thin films (filled red circles) and bulk crystals reported in Ref.\cite{Shu11} (open blue circles). (b) Temperature dependence of the resistivity at low temperatures. The dased lines are the extrapolation above $T_c$.  }
\hspace*{3mm}
\end{center}
\end{figure}

Next we discuss the effect of Yb-substitution on the normal state transport properties.  As shown in Fig.~2a, a broad maximum of $\rho(T)$ associated with the formation of  Kondo coherence is observed at $T_{\rm coh}(x)$ in Yb-substituted CeCoIn$_5$.    Figure~3a depicts the $x$-dependence of $T_{\rm coh}$, which is simply defined as the peak position of $\rho(T)$.     Yb-substitution  seriously reduces  the Kondo coherence and $T_{\rm coh}$ appears to go to zero at around $x=0.5$.   Since it is highly unlikely that Fermi temperature is dramatically suppressed by Yb-substitution, the observed reduction of $T_{\rm coh}$ arises from the destruction of Kondo coherence by the Yb$^{2+}$Kondo-holes in the Ce$^{3+}$$f$-electron lattice. In Yb-doped compounds $\rho(T)$ exhibit sub-$T$-linear behavior below $T_{\rm coh}$.  We try to fit the resistivity by a power law $\rho(T)=\rho_0+AT^\alpha$ (Fig.~3b).    The resistivity can be fitted very well by this equation for $x$~=~0 and  0.02 with the residual resistivity $\rho_0$= 4.8 and 6.0  $\mu \Omega$~cm and $\alpha$=0.88 and  0.80, respectively,  in a wide temperature range below $\sim T_{\rm coh}/2$ down to $T_c$.   For $x\geq$~0.05,  we cannot fit $\rho(T)$  by this equation in a wide temperature range and we then estimate $\rho_0$ by the polynomial fitting above $T_c$ to $T_{\rm coh}/2$ (see the dashed lines in Fig~3b).   As shown in Fig.~4a, the residual resistivity nearly linearly increases with $x$.    The present results are very different from those reported in bulk single crystals, in which $T_{\rm coh}$ is nearly $x$-independent as shown in open symbols in Fig.~3a and $\alpha$ barely changes with $x$ up to $x$~=~0.1 \cite{Shu11}.

The measurements of Hall effect provide vital information for the nature of the Yb-ion in the Ce-Kondo lattice.  As shown in Fig.~2b, the Hall effect is very sensitive to the Yb-substitution.    At $T>T_{\rm coh}$,  the absolute value of  $R_{\rm H}$ is small and nearly temperature independent for all $x$.   This  is consistent with the the Vegard's law which suggests divalent Yb in trivalent Ce lattice.   Below $T_{\rm coh}$, the enhancement of  $|R_{\rm H}|$ is strongly suppressed by the Yb-doping. The low-temperature enhancement of $| R_{\rm H} |$, which is significantly larger than $|1/ne|$ where $n$ is the carrier concentration, have also been reported in other strongly correlated electron systems including cuprates \cite{Hwa94}  iron-pnictides \cite{Kasa10}, $\kappa$-(BEDT-TTF)$_2$Cu(NCS)$_2$ \cite{Sus97} and V$_{2-y}$O$_3$ \cite{Ros98}.    It has been shown that the striking enhancement of $|R_{\rm H}|$ at low temperatures can be accounted for in terms of  \textquotedblleft hot spot\textquotedblright on the Fermi surface (FS) and backflow effect, both of which originate  from strong AF fluctuations \cite{Kon99}.   The hot spot is a FS region where the electron lifetime is unusually short, which appears at the positions where AF Brillouin zone boundary intersects with the FS.   Since the hot spot  does not contribute to the electron transport,  the effective carrier concentration is reduced, which results in the enhancement of $|R_{\rm H}|$ from $|1/ne|$ \cite{Pines97}.  However, this effect is not enough to explain the dramatic enhancement of $|R_{\rm H}|$ below $T_{\rm coh}$. Another important effect is the backflow effect accompanied by the anisotropic scattering, which is called the current vertex correction \cite{Kon99}.  Backflow is a polarized current caused by a quasiparticle excitation inherent in Fermi liquid.   In the presence of the backflow effect, the total current is not parallel to the Fermi velocity and then the Hall coefficient  is seriously modified from the Boltzmann value derived from the curvature of the FS.   It has been shown that the backflow effect can largely enhance $|R_{\rm H}|$ in Ce$M$In$_5$ ($M$=Co, Rh and Ir) \cite{Nakajima07,Nakajima08}.  

The observed strong suppression of $|R_{\rm H}|$ by Yb-substitution indicates that  the backflow effect is seriously reduced.   There are two possible origins for this.  The first is the reduction of AF fluctuations and the second is the increase of isotropic impurity scattering by nonmagnetic Yb-ions.    In the former case, since the system is tuned away from the quantum critical point, the $T$-linear resistivity is expected to change to $T^2$-dependence accompanied by a reduction of $\rho_0$ \cite{Nakajima07}.  However, such a trend is not observed here.  Instead, the $T$-linear resistivity changes to sub-$T$-linear and $\rho_0$ increases with Yb-substitution.  This increase of $\rho_0$  is consistent with the increase of isotropic scattering.  In addition, it has been shown that in the presence of strong AF fluctuations, sub-$T$-linear dependence of $\rho(T)$ appears with increasing impurity scattering \cite{Kon06}, which is again consistent with the present results.  Thus the normal state transport properties suggest that Yb-ions are divalent and act as impurity scattering centers, rather than a suppressor of AF fluctuations.   

\begin{figure}[t]
\begin{center}
\hspace*{-3mm}
\includegraphics[width=1.07\linewidth]{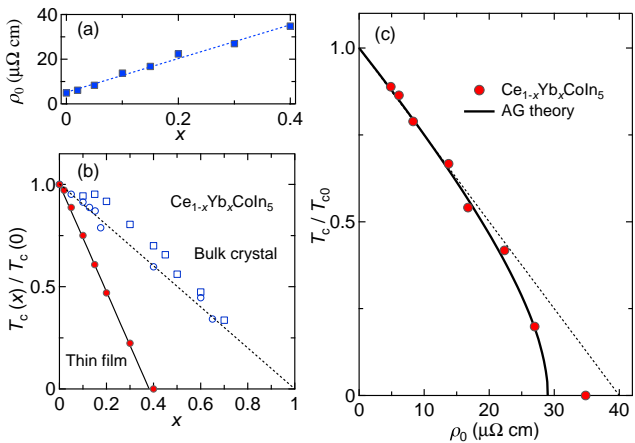}
\caption{(color online). (a) $x$-dependence of the residual resistivity.  Dotted line is a guide for eyes.  (b) $x$-dependence of the superconducting transition temperature normalized by the value at $x$=0 for thin films (filled red circles) and bulk crystals (open blue squares \cite{Shu11} and open circles \cite{Boo11}).   (c) $T_c/T_{c0}$ plotted as a function of residual resistivity.  Dotted line is the linear extrapolation from the low $\rho_0$ region.  The solid curve represents the result fitted by Abrikosov-Gorkov (AG) theory.}
\hspace*{3mm}
\end{center}
\end{figure}

Next we discuss the influence of Yb-substitution on the superconducting properties. As shown in Fig.~3b, sharp resistive transitions are observed for $x\leq0.20$, but the transition is slightly broader for $x$~=~0.30.  No superconducting transition is observed for $x$~=~0.40 down to 0.3~K.   Figure~4b displays the $x$-dependence of $T_c$, which is defined as the midpoint of the resistive transition,  normalized by $T_c$ of $x$=0.  $T_c$ decreases linearly with $x$ and goes to zero at around $x=0.40$.  We stress that the observed Yb-doping evolution of $T_c$ is again in marked contrast to that reported in bulk single crystals shown by open symbols in Fig.~4b.   In Fig.~4c we plot $T_c/T_{c0}$ (see below) against $\rho_0$.  In the low $\rho_0$ range,  $T_c/T_{c0}$  decreases linearly as shown by the dotted line.  When $\rho_0$ exceeds $\sim 18~\mu \Omega$~cm, $T_c$ begins to deviate from  linearity.   We analyze this trend of $T_c/T_{c0}$ in accordance with the AG theory of nonmagnetic impurity effect in $d$-wave superconductors \cite{Maki95}.     The fitting parameters used to compare experiment to theory are  the slope in the low $\rho_0$ region and  $T_{c0}$, where $T_{c0}$ is the transition temperature with no pair-breaking.  The solid line in Fig.~4c is the result of fitting obtained by using $T_{c0}=2.20$~K and the initial slope -0.025~$(\mu\Omega~{\rm cm})^{-1}$.   As shown in Fig.~4c, the suppression of $T_c$ is well reproduced by the AG pair breaking curve in the whole $\rho_0$ range.   

The fact that our data can be well described by the AG theory immediately implies that the Yb-ions are randomly distributed with weak inter-ion correlation. Moreover,  the \textquotedblleft Swiss cheese\textquotedblright model \cite{Bau11} is inadequate because such a model will give a $T_c(\rho_0)$-curve that lies above  the dotted line for high $\rho_0$ in Fig.~4c \cite{Fra97}.   The effect of Yb-substitution on CeCoIn$_5$ bears striking resemblance to  other rare earth substitutions, although Yb is divalent while other rare earths are trivalent.  This suggests that these ions act as  impurity centers with unitary scattering, regardless of their valence.   Furthermore, the  valence fluctuations of Yb-ions, if present at all, do not appear to play an important role on the physical properties in the normal state as well as the superconducting state.  At the present stage, the essential difference of the Yb-substitution effect between thin films and bulk single crystals remains an open question.   A possible origin for this may be that  bulk crystals contain some regions where inter-Yb-ion correlation is important \cite{Dzero12}.   Further studies are  required to clarify this issue.

To summarize,  in  high quality Ce$_{1-x}$Yb$_x$CoIn$_5$ epitaxial thin films, Yb-ions are divalent and no  signature of the valence fluctuation is observed.  The Yb-substitution leads to a strong suppression of  the superconductivity and Kondo coherence.   The suppression of $T_c$ can be well described by AG theory.   These results indicate that Kondo-holes created by Yb-ions act as nonmagnetic impurity scatters in the  Kondo lattice with no serious reduction of AF fluctuations.    These are in sharp contrast to previous studies performed using bulk single crystals, which claim the importance of valence fluctuations of Yb-ions.  The present work also emphasizes the uniqueness of the epitaxial films in the study of the impurity effect on the  $f$-electron Kondo lattices,  due to the prospect of preparing highly homogeneous doped systems.

We thank C.~Capan, H.~Ikeda, N.~Kawakami, H.~Kontani, M.~B.~Maple and L.~Shu for helpful discussions. This work was supported by Grant-in-Aid for the Global COE program ``The Next Generation of Physics, Spun from  Universality and Emergence", Grant-in-Aid for Scientific Research on Innovative Areas ``Heavy Electrons'' (No. 20102006, 23102713) from MEXT of Japan, and KAKENHI from JSPS.


\end{document}